\documentstyle[epsfig]{mn}

\newif\ifAMStwofonts

\ifoldfss
  \newcommand{\rmn}[1] {{\rm #1}}

  \ifCUPmtlplainloaded \else
    \NewTextAlphabet{textbfit} {cmbxti10} {}
    \NewTextAlphabet{textbfss} {cmssbx10} {}
    \NewMathAlphabet{mathbfit} {cmbxti10} {} 
    \NewMathAlphabet{mathbfss} {cmssbx10} {} 
  \fi
  \ifAMStwofonts
    \ifCUPmtlplainloaded \else
      \NewSymbolFont{upmath} {eurm10}
      \NewSymbolFont{AMSa} {msam10}
      \NewMathSymbol{\upi}     {0}{upmath}{19}
      \NewMathSymbol{\umu}     {0}{upmath}{16}
      \NewMathSymbol{\upartial}{0}{upmath}{40}
      \NewMathSymbol{\leqslant}{3}{AMSa}{36}
      \NewMathSymbol{\geqslant}{3}{AMSa}{3E}

      \let\geq=\geqslant \let\ge=\geqslant
    \fi
  \fi
\fi 

\ifnfssone
  \newmathalphabet{\mathit}
  \addtoversion{normal}{\mathit}{cmr}{m}{it}
  \addtoversion{bold}{\mathit}{cmr}{bx}{it}
  \newcommand{\rmn}[1] {\mathrm{#1}}

  \newmathalphabet{\mathbfit} 
  \addtoversion{normal}{\mathbfit}{cmr}{bx}{it}
  \addtoversion{bold}{\mathbfit}{cmr}{bx}{it}
  \newmathalphabet{\mathbfss} 
  \addtoversion{normal}{\mathbfss}{cmss}{bx}{n}
  \addtoversion{bold}{\mathbfss}{cmss}{bx}{n}
  \ifAMStwofonts
    \ifCUPmtlplainloaded \else
      \UseAMStwoboldmath
      \makeatletter
      \new@mathgroup\upmath@group
      \define@mathgroup\mv@normal\upmath@group{eur}{m}{n}
      \define@mathgroup\mv@bold\upmath@group{eur}{b}{n}
      \edef\UPM{\hexnumber\upmath@group}
      \new@mathgroup\amsa@group
      \define@mathgroup\mv@normal\amsa@group{msa}{m}{n}
      \define@mathgroup\mv@bold\amsa@group{msa}{m}{n}
      \edef\AMSa{\hexnumber\amsa@group}
      \makeatother
      \mathchardef\upi="0\UPM19
      \mathchardef\umu="0\UPM16
      \mathchardef\upartial="0\UPM40
      \mathchardef\leqslant="3\AMSa36
      \mathchardef\geqslant="3\AMSa3E

      \let\geq=\geqslant \let\ge=\geqslant
    \fi
  \fi
\fi 

\ifnfsstwo
  \newcommand{\rmn}[1] {\mathrm{#1}}

  \DeclareMathAlphabet{\mathbfit}{OT1}{cmr}{bx}{it}
  \SetMathAlphabet\mathbfit{bold}{OT1}{cmr}{bx}{it}
  \DeclareMathAlphabet{\mathbfss}{OT1}{cmss}{bx}{n}
  \SetMathAlphabet\mathbfss{bold}{OT1}{cmss}{bx}{n}
  \ifAMStwofonts
    \ifCUPmtlplainloaded \else
      \DeclareSymbolFont{UPM}{U}{eur}{m}{n}
      \SetSymbolFont{UPM}{bold}{U}{eur}{b}{n}
      \DeclareSymbolFont{AMSa}{U}{msa}{m}{n}
      \DeclareMathSymbol{\upi}{0}{UPM}{"19}
      \DeclareMathSymbol{\umu}{0}{UPM}{"16}
      \DeclareMathSymbol{\upartial}{0}{UPM}{"40}
      \DeclareMathSymbol{\leqslant}{3}{AMSa}{"36}
      \DeclareMathSymbol{\geqslant}{3}{AMSa}{"3E}

      \let\geq=\geqslant \let\ge=\geqslant
    \fi
  \fi
\fi 

\ifCUPmtlplainloaded \else
  \ifAMStwofonts \else 
    \def\upi{\pi}
    \def\umu{\mu}
    \def\upartial{\partial}
  \fi
\fi

\title{The First Galaxies:  Signatures of the Initial Starburst}
\author[J. L. Johnson, et al.]
       {Jarrett L. Johnson$^1$$^,$$^3$$^,$$^4$\thanks{E-mail: jljohnson@astro.as.utexas.edu}, Thomas H. Greif $^2$$^,$$^5$, 
          Volker Bromm$^1$$^,$$^3$, \newauthor Ralf S. Klessen$^2$ and Joseph Ippolito$^1$\\
 $^1$Department of Astronomy, University of Texas, Austin, TX 78712, USA \\
 $^2$Zentrum f{\"u}r Astronomie der Universit{\"a}t Heidelberg,  Institut f{\"u}r Theoretische Astrophysik, Albert-Ueberle-Str. 2, 69120 Heidelberg, Germany \\
 $^3$Texas Cosmology Center, University of Texas at Austin, TX 78712 \\
 $^4$Wendell Gordon Fellow\\
 $^5$Fellow of the International Max Planck Research School for Astronomy and Cosmic Physics at the University of Heidelberg\\}

\pubyear{2009}

\begin{document}

\maketitle
\topmargin-1cm

\label{firstpage}

\begin{abstract}
Detection of the radiation emitted from the first galaxies at $z$ $\geq$ 10 will be made possible in the
next decade, with the launch of the {\it James Webb Space Telescope} (JWST). We
carry out cosmological radiation hydrodynamics simulations of
Population III (Pop III) starbursts in a 10$^8$ ${\rmn M}_{\odot}$ dwarf galaxy at $z$ $\sim$ 12.5.  For different
star formation efficiencies and stellar initial mass functions (IMFs), we calculate
the luminosities and equivalent widths (EWs) of the recombination lines H$\alpha$, Ly$\alpha$, and He~{\sc ii} $\lambda$1640, 
under the simple assumption that the stellar population does not evolve over the first $\sim$ 3 Myr of the starburst.  Although only $\la$ 40 percent of 
the gas in the central 100 pc of the galaxy is photoionized, we find that photoheating by massive stars causes a strong 
dynamical response, which results in a
weak correlation between luminosity emitted in hydrogen recombination lines and the total mass in stars. 
However, owing to the low escape fraction of He~{\sc ii}-ionizing photons, the luminosity emitted in   
He~{\sc ii} $\lambda$1640 is much more strongly correlated with the total stellar mass.
The ratio of the luminosity in He~{\sc ii} $\lambda$1640 to that in Ly$\alpha$ or H$\alpha$ is a complex function of the density field 
and the star formation rate, but is found to be a good indicator of the IMF in many cases.  The ratio of observable fluxes 
is F$_{\rm \lambda 1640}$/F$_{\rm H\alpha}$ $\sim$ 1 for clusters of 100 ${\rmn M}_{\odot}$ Pop~III stars and 
F$_{\rm \lambda 1640}$/F$_{\rm H\alpha}$ $\sim$ 0.1 for clusters of 25 ${\rmn M}_{\odot}$ Pop~III stars.        
The EW of the He~{\sc ii} $\lambda$1640 emission line is the most reliable IMF indicator, its value varying between 
$\sim$ 20 and $\sim 200 {\rm\AA}$  for a massive and very massive Pop~III IMF, respectively.  
Even the bright, initial stages of Pop~III starbursts in the first dwarf galaxies will likely not be directly detectable by the JWST, 
except in cases where the flux from these galaxies is strongly magnified through gravtitational lensing.
Instead, the JWST may discover already more massive, and hence more chemically evolved, galaxies in which primordial 
star formation has largely ceased, or is contaminated with more normal, Pop~I/II, star formation.

\end{abstract}


\section {Introduction}
The epoch of the first galaxies marked a fundamental transition in the Universe, ending the Cosmic Dark Ages, 
beginning the process of reionization, and witnessing the rapid proliferation of star formation and black hole 
growth (e.g. Barkana \& Loeb 2001; Bromm \& Larson 2004).  While the theory of primordial star formation and 
early galaxy formation has rapidly developed (see Ciardi \& Ferrara 2005; Glover 2005; Barkana \& Loeb 2007), observations of the 
first galaxies at redshifts z $\ga$ 10 have so far been out of reach (but see Stark et al. 2007).  In the coming 
decade, the {\it James Webb Space Telescope} (JWST) promises to provide direct observations of this critical 
period in cosmic history, allowing to place new constraints on the stellar initial mass function (IMF) at high redshift, 
on the luminosity function of the first galaxies, and on the progress of the early stages of reionization 
(e.g. Barton et al. 2004; Gardner et al. 2006; Windhorst et al. 2006; Ricotti et al. 2008; Haiman 2008).

The IMF of the stellar populations which form in the first galaxies is of 
central importance in determining their properties and impact on early cosmic evolution. 
The current theoretical consensus posits that the first 
stars, which formed in isolation in dark matter (DM) minihaloes, likely had masses of the order of 100 
${\rmn M}_{\odot}$ (Bromm et al. 1999, 2002; Abel et al. 2002; Yoshida et al. 2006, 2008; McKee \& Tan 2008).  
In the first galaxies, which form in DM haloes with masses $\ga$ 10$^8$ ${\rmn M}_{\odot}$ (Greif et al. 2008; 
Wise et al. 2008), there is no such theoretical consensus on the IMF, as the initial conditions of the star-forming gas are uncertain (see Jappsen et al. 2009a,b). 
A large fraction of these first 
galaxies are likely to already host Population II (Pop II) star formation, due to previous metal enrichment (Johnson et al. 2008; see also Trenti \& Stiavelli 2009; 
Clark et al. 2008).  
However, clusters of primordial stars likely form in some fraction of the first galaxies, owing to either 
inhomogeneities in the Lyman-Werner (LW) background which can suppress star formation where galaxies are strongly clustered 
(Dijkstra et al. 2008; see also Ahn et al. 2008) or to the direct collapse of the 
first stars to black holes, thereby locking up the metals produced in their cores (e.g. Heger et al. 2003).  

Due to the hard spectra of massive metal-free stars, strong nebular emission in helium recombination lines has been suggested as an 
observable indicator 
of a population of such stars (e.g. Bromm et al. 2001; Oh et al. 2001; Tumlinson et al. 2001; Schaerer 2002).  In particular, a high 
ratio of the 
luminosity emitted in He~{\sc ii} $\lambda$1640 to that 
emitted in Ly$\alpha$ or H$\alpha$ may be a signature of a galaxy hosting massive Pop~III star formation, and has already served as the 
basis for searches for such 
galaxies (e.g. Dawson et al. 2004; Nagao et al. 2005, 2008).  In addition, high equivalent widths (EWs) of Ly$\alpha$ and He~{\sc ii} 
$\lambda$1640 are expected to characterize 
galaxies undergoing a Pop~III starburst (e.g. Schaerer 2003).  While no definitive detections of such galaxies have been achieved to 
date, observations 
of galaxies at 3 $\la$ $z$ $\la$ 6.5 with large Ly$\alpha$ EW and strong He~{\sc ii} $\lambda$1640 emission may indicate that some 
galaxies host Pop~III star formation even at such relatively low redshift (see Jimenez \& Haiman 2006; Dijkstra \& Wyithe 2007).

Previous analytical calculations of the recombination radiation expected from the first galaxies 
have been carried out under a number of idealized assumptions, namely of a static, uniform density 
field, and of the formation of a static Str{\"o}mgren sphere.  Taking a complementary approach, we study here the properties
of the recombination radiation emitted by the first galaxies with a focus on how the dynamical evolution of the galaxy
affects the properties of this radiation. We present high-resolution cosmological radiation hydrodynamics 
simulations of the production of nebular emission from a cluster of primordial stars formed within the first 
galaxies.  We resolve the H~{\sc ii} and He~{\sc iii} regions generated by the stellar cluster, thereby 
arriving at improved predictions for the emission properties of the first galaxies, which will be tested by the JWST.

Our paper is organized as follows. In Section 2, we describe our simulations and the methods used in their analysis; 
in Section 3 our results are presented, along with the implications for both IMF and star formation rate (SFR) indicators; 
in the final Section 4, we summarize and give our conclusions.             

\begin{figure}
\vspace{2pt}
\epsfig{file=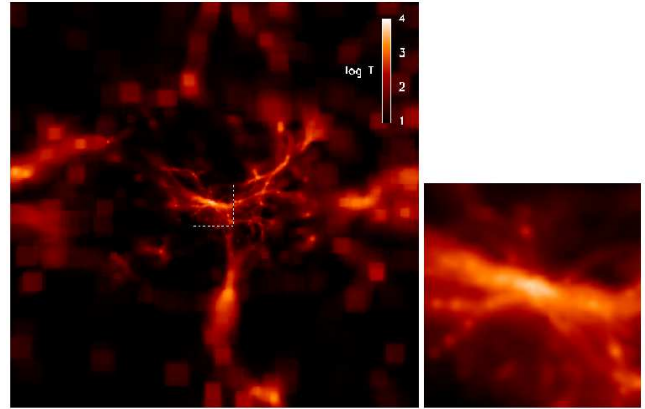,width=8.5cm,height=5.5cm}
\caption{The projected gas temperature at $z$ = 12.7, just before the stellar cluster turns on.  Shown at left is the entire 1.4 
Mpc (comoving) simulation box.  The region of highest resolution is
in the center of our multi-grid simulation box; magnified at right is the central $\sim$ 10 kpc (physical),  where the $\sim$ 10$^8$ 
${\rmn M}_{\odot}$ dark matter halo hosting the stellar cluster is located. Note that the gas is heated to 
$\ge$ 10$^4$~K when passing through the virialization shock.}
\end{figure}

\section{Methodology}
We carry out radiation hydrodynamics simulations which track the impact of the radiation from Pop~III stellar clusters forming
promptly within a dwarf galaxy at $z$ $\sim$ 12.5.   
In this section we describe the simulations and the calculations carried out in analyzing their outcomes. 

\subsection {The simulations}

As with previous work, for our three-dimensional numerical simulations we employ the parallel version of GADGET 
(version 1), which includes a tree (hierarchical) gravity solver combined with the smoothed particle hydrodynamics 
(SPH) method for tracking the evolution of gas (Springel et al. 2001; Springel \& Hernquist 2002).  Along with 
H$_2$, H$_2^{+}$, H, H$^-$, H$^+$, e$^-$, He, He$^{+}$, and He$^{++}$, we have included the five deuterium species 
D, D$^+$, D$^-$, HD and HD$^+$, using the same chemical network as in Johnson \& Bromm (2006, 2007).

For our simulation of the assembly of a dwarf galaxy at $z$ $\sim$ 12.5, we have employed multi-grid initial 
conditions which offer higher resolution in the region where the galaxy forms (e.g. Kawata \& Gibson 2003).
We initialize the simulation according to the $\Lambda$CDM power spectrum at $z$ = 100, adopting the cosmological 
parameters $\Omega_{m}=1 - \Omega_{\Lambda}=0.3$, $\Omega_{B}=0.045$, $h=0.7$, and $\sigma_{8}=0.9$, close to the 
values measured by the {\it Wilkinson Microwave Anisotropy Probe} in its first year (Spergel et al. 2003). Here we use a periodic box with a comoving 
size $L$ = 1 $h^{-1}$ Mpc for the parent grid.  Our simulations use $N_{\rm DM}$ = $N_{\rm SPH}$ 
= 1.05 $\times$ 10$^6$ particles for DM and gas, where the SPH particle mass is $m_{\rm SPH}$ $\sim$  120 ${\rm M}_{\odot}$ in the region 
with the highest resolution. For further details on the technique employed to generate our multi-grid initial 
conditions, see Greif et al. (2008).  The maximum gas density that we resolve is $n_{\rm res}$ $\sim$ 10$^3$ cm$^{-3}$, 
while gas at higher densities is accreted onto sink particles, as described in Johnson et al. (2007).  We have also 
included the effect of a LW background radiation field, at a level of $J_{\rm LW}$ = 0.04 $\times$ 
10$^{-21}$ erg s$^{-1}$ cm$^{-2}$ Hz$^{-1}$ sr$^{-1}$ , just as in Johnson et al. (2008). 

To capture the effects of the ionizing radiation emitted by a single Pop~III stellar cluster within the dwarf galaxy, we approximate
the cluster as a point source located at the center of the most massive DM halo in our simulation box at $z$ = 12.7. The 
projected gas temperature in the entire simulation box at this redshift is shown in Fig. 1 (left panel); the right panel
shows the temperature in the region of the host halo at the center of the box, which has a virial mass of 9 $\times$ 
10$^7$ ${\rm M}_{\odot}$, characteristic of the first galaxies.  At each timestep, we find the boundaries of both the H~{\sc ii} and He~{\sc iii} regions generated by the stellar cluster 
using a ray-tracing technique that improves our earlier implementation (Johnson et al. 2007). 

The procedure used to calculate the Str\"omgren sphere around the stellar cluster for a given time-step 
$\Delta t$ is similar to the ray-tracing scheme used in Johnson et al. (2007).  We create a spherical 
grid centered at the location of the cluster, consisting of $\sim$ 1.2 $\times$ 10$^4$ rays and $1000$ linearly 
spaced radial bins.  We resolve the central kiloparsec around the source, roughly the virial radius of 
the host halo, with 250 radial bins, while the remaining 750 bins are linearly spaced out to $\sim$ 20 kpc.

In a single, parallelized loop, the Cartesian coordinates of all particles are converted to spherical 
coordinates, such that their density and chemical abundances may be mapped to the bins corresponding to 
their radius, azimuth and zenith, denoted by $r$, $\theta$ and $\phi$, respectively. 
To avoid missing dense clumps, particles contribute to bins independent of distance, but proportional 
to their density squared.  Once this preliminary step is complete, it is straightforward to solve the 
ionization front equation along each ray:

\begin{equation}
n_{n}r_{\rm I}^{2}\frac{{\rm d}r_{\rm{I}}}{{\rm d}t}=\frac{{Q}_{\rm{ion}}}{4\pi}-\alpha_{\rm{B}}\int_{0}^{r_{\rm{I}}}n_{e}n_{+}r^{2}{\rm d}r\mbox{\ ,}
\end{equation}
where $r_{\rm{I}}$ denotes the position of the ionization front, ${Q}_{\rm{ion}}$ the number of 
ionizing photons emitted per second by the stellar cluster, 
$\alpha_{\rm{B}}$ the case B recombination coefficient, and $n_{n}$, $n_{e}$ and $n_{+}$ the number 
densities of nonionized particles, electrons and ionized particles, respectively. The numbers of 
H~{\sc i}-, He~{\sc i}- and He~{\sc ii}-ionizing photons are Q$_{\rm ion}$ = $N_{\rm *}$$q_{\rm ion}$, where 
$N_{\rm *}$ is the number of stars in the cluster (here we assume that all have the same mass) 
and $q_{\rm ion}$ is the number of ionizing photons emitted by a single star, given by 

\begin{equation}
{q}_{\rm{ion}}=\frac{\pi L_{*}}{\sigma T_{\rm{eff}}^{4}}\int_{\nu_{\rm{min}}}^{\infty}\frac{B_{\nu}}{h\nu}\rm{d}\nu\mbox{\ ,}
\end{equation}
where $\sigma$ is the Stefan-Boltzmann constant, $\nu_{\rm{min}}$ denotes the minimum frequency corresponding to the 
ionization thresholds of H~{\sc i}, He~{\sc i} and He~{\sc ii}, and we assume that massive Pop~III stars emit a 
blackbody spectrum $B_{\nu}$ (in $\rm{erg}~\rm{s}^{-1}~\rm{cm}^{-2}~\rm{Hz}^{-1}~\rm{sr}^{-1}$) 
with an effective temperature $T_{\rm{eff}}$ and a luminosity $L_{*}$ (e.g. Schaerer 2002).

To obtain a discretization of the ionization front equation, we replace the integral on the right-hand side of equation~(1) by a discrete sum:
\begin{equation}
\int_{0}^{r_{\rm{I}}}n_{e}n_{+}r^{2}{\rm d}r=\sum_{i}n_{e,i}n_{+,i}r_{i}^{2}\Delta r\mbox{\ ,}
\end{equation}
where $\Delta r$ is the radial extent of the individual bins.  Similarly, the left-hand side of equation~(1), which models the propagation of the ionization front into neutral gas, is discretized by
\begin{equation}
n_{n}r_{\rm I}^{2}\frac{{\rm d}r_{\rm{I}}}{{\rm d}t}=\frac{1}{\Delta t}\sum_{i}n_{n,i}r_{i}^{2}\Delta r_{i}\mbox{\ ,}
\end{equation}
where $\Delta t$ is the current time-step and the summation is over radial bins starting with the bin lying immediately outside of $r_{\rm{I,old}}$, the position of the ionization front at the end of the previous time-step, and ending with the bin lying at the new position of the ionization front. We perform the above steps separately for the H~{\sc ii} and He~{\sc iii} regions, since they require distinct heating and ionization rates. 

We have chosen the size of the bins that are used in our ray-tracing routine to roughly match the volume 
of gas represented by a single SPH particle within the $\sim$ 1 kpc virial radius of the halo hosting the stellar cluster, 
such that the boundaries of the photoionized regions are 
maximally resolved while also reliably conserving ionizing photons.  However, in some cases it may occur that the mass contained in a bin is smaller 
than that of the SPH particle contained 
within it, such that ionizing the entire SPH particle involves ionizing more gas than is contained in the bin.  In turn, this can lead to 
an overestimate of the number of recombinations.  While this effect is minor in our simulations, in the calculations presented 
below we enforce that the total number of recombinations does not exceed the total number of ionizing photons available.

We carry out four simulations, each with a different combination of IMF and total cluster mass. 
For the IMF, we assume for simplicity, and in light of the still complete uncertainty regarding its detailed shape, that the cluster
consists either entirely of
25 ${\rmn M}_{\odot}$ or 100 ${\rmn M}_{\odot}$ Pop~III stars. These choices are meant to bracket the 
expected characteristic mass for Pop~III stars formed in the first galaxies, which depending on the cooling properties of 
the gas may be Pop~III.2 stars with masses of order 10 ${\rmn M}_{\odot}$ or, possibly, Pop~III.1 stars with masses perhaps 
an order of magnitude higher (Johnson \& Bromm 2006; Greif et al. 2008; McKee \& Tan 2008; but see Jappsen et al. 2009a). 
For each of these IMFs, we vary the total stellar mass in the cluster, choosing either 2.5 $\times$ 10$^3$ ${\rmn M}_{\odot}$ or 2.5 $\times$ 
10$^4$ ${\rmn M}_{\odot}$ for the total mass in stars.  These choices correspond to $\sim$ 1 and $\sim$ 10 percent, respectively, 
of the cold gas available for star formation within the central few parsecs of such a primordial dwarf galaxy (see Wise et al. 2008; Regan \& Haehnelt 2009).    
We calculate the ionizing flux from each of these clusters, assuming blackbody stellar spectra at 7 $\times$ 10$^4$ and 10$^5$
K and bolometric luminosities of 6 $\times$ 10$^4$ and 10$^6$ ${\rmn L}_{\odot}$, for the 25 and 100 ${\rmn M}_{\odot}$ stars, respectively, appropriate for 
metal-free stars on the main sequence (Marigo et al. 2001).  

For simplicity, we have chosen to keep the input stellar spectra constant in time over the course of
the simulations.  Accordingly, we run the simulations only for 3 Myr, which is roughly the hydrogen-burning 
timescale of a 100 ${\rmn M}_{\odot}$ primordial star, and about half that of a 25 ${\rmn M}_{\odot}$ 
primordial star.  We note that while the H~{\sc i}-ionizing flux from 100 ${\rmn M}_{\odot}$ primordial stars is roughly constant 
over this timescale, the He~{\sc ii}-ionizing flux decreases by a factor of $\sim$ 2 by a stellar age of 
2 Myr, and by a much larger factor near the end of hydrogen-burning as the star evolves to the red 
(e.g. Marigo et al. 2001; Schaerer 2002).  We note, however, that stellar models accounting 
for the effects of rotation yield less precipitous drops in the emitted ionizing flux with time, as fast rotation, especially of 
low-metallicity stars, can keep the stars on bluer evolutionary tracks (e.g. Yoon \& Langer 2005; Woosley \& Heger 2006; 
V\'azquez et al. 2007); indeed, Pop~III stars may have been fast rotators (see Chiappini et al. 2008).  
Nonetheless, the results that we derive pertaining 
to He~{\sc ii} recombination emission from clusters of 100 ${\rmn M}_{\odot}$ stars may be, strictly speaking, 
only reliable for stellar ages $\la$ 2 Myr. 
An in-depth study of the impact that stellar evolution has on the emission properties
of primordial galaxies is given in Schaerer (2002); in the present work, we take a 
complementary approach and instead focus on how the emission properties are affected by the hydrodynamic 
evolution of the gas in the first galaxies.  

We make the related simplifying assumption that the stellar cluster forms instantaneously.  
This is valid if the timescale for star formation $t_{\rm SF}$ is much shorter than
the lifetime of the stars that we consider, or $t_{\rm SF}$ $\ll$ 3 Myr.  If we assume that stars form on the order of 
the free-fall time $t_{\rm ff}$, and take it that the star cluster forms within the central $\sim$ 1 pc of 
the halo (see e.g. Wise et al.2008; Regan \& Haehnelt 2009), then we find $t_{\rm SF}$ $\sim$ 5 $\times$ 10$^5$ yr,
for which our assumption is marginally valid.  We note that more work is needed to accurately determine the star formation 
timescale in the first galaxies, as the works cited here neglect, in particular, the important effect of molecular cooling 
on the evolution of the primordial gas.

\subsection{Deriving the observational signature}
The simulations described above allow us to calculate the luminosities and equivalent widths 
of the recombination lines emitted from high-redshift dwarf galaxies during a primordial starburst.  
A related quantity we obtain is the escape fraction of ionizing photons from such a galaxy.  
Here we describe each of these calculations.   

\subsubsection{Escape fraction of ionizing photons}
Photons which escape the host halo from which they are emitted proceed to reionize the intergalactic medium (IGM), where densities are generally 
very low, yielding long recombination times.  Ionizing photons which do not escape the host halo 
are, however, available to ionize dense gas which recombines quickly, leading to appreciable emission in recombination lines.
Therefore, the luminosity of a galaxy in recombination radiation is intimately related to the escape fraction of ionizing photons.
The escape fraction of ionizing photons from the halo hosting the stellar cluster is 
given by subtracting the number of recombinations $Q_{\rm rec}$ per second within the virial radius from the total number of ionizing photons emitted by the cluster:

\begin{equation}
f_{\rm esc} = \left(Q_{\rm ion} - Q_{\rm rec}\right) \left(Q_{\rm ion}\right)^{-1}\mbox{\ ,}
\end{equation}
again with $Q_{\rm ion}$ = $N_{\rm *}$$q_{\rm ion}$.  This equation is valid under the assumption that 
within the host halo the number of ionizing photons which fail to escape is balanced by the number of recombinations within 
the halo.  This is a reasonable assumption, since the number of atoms 
which become ionized within the host halo is far less than the total 
number of recombinations that occur in the halo, the ionization of 
previously neutral gas being the only other sink for ionizing photons 
within the halo.
The number of recombinations is given as

\begin{equation}
Q_{\rm rec} = \sum_{\rm i} \alpha_{\rm B} \frac{m_{\rm i}}{\rho_{\rm i}} \left[\frac{\rho_{\rm i}}{\mu_{\rm i} m_{\rm H}}\right]^2 f_{\rm e} f_{\rm H II}  \mbox{\ ,}
\end{equation}
where $\alpha_{\rm B}$ is the case~B recombination coefficient for hydrogen,
$m_{\rm H}$ the mass of a hydrogen atom, while
$m_{\rm i}$, $\mu_{\rm i}$ and $\rho_{\rm i}$ are the 
total mass, mean molecular weight, and mass density of the $i$th SPH particle, respectively. For each SPH particle, $f_{\rm HII}$ and $f_{\rm e}$
denote the fraction of nuclei in H~{\sc ii} and the fraction of free electrons, respectively.

Here the summation is over all SPH particles within the virial radius of the host halo, or within a physical distance 
of $\sim$ 1~kpc from the central stellar cluster. 
We calculate the escape fractions of both H~{\sc i}-ionizing and He~{\sc ii}-ionizing photons.  These quantities are generally not equal, 
and they each contribute to determining the radiative signature of the initial starbursts in the first galaxies.  

\subsubsection{Luminosity in recombination lines}
For each of our simulations, we compute the luminosity emitted from photoionized regions in each of three recombination 
lines:  H$\alpha$, Ly$\alpha$, and He~{\sc ii} $\lambda$1640.  
These luminosities are calculated by again summing up the contributions from all SPH particles within the virial radius, where
virtually all of the recombination line luminosity emerges, as follows:

\begin{equation}
L_{\rm H\alpha} = \sum_{\rm i} j_{\rm H \alpha} \frac{m_{\rm i}}{\rho_{\rm i}} \left[\frac{\rho_{\rm i}}{\mu_{\rm i} m_{\rm H}}\right]^2 f_{\rm e} f_{\rm H II}\mbox{\ ,}
\end{equation}

\begin{equation}
L_{\rm Ly\alpha} = \sum_{\rm i} j_{\rm Ly \alpha} \frac{m_{\rm i}}{\rho_{\rm i}} \left[\frac{\rho_{\rm i}}{\mu_{\rm i} m_{\rm H}}\right]^2 f_{\rm e} f_{\rm H II}\mbox{\ }
\end{equation}

\begin{equation}
L_{\rm \lambda 1640} = \sum_{\rm i} j_{\rm \lambda 1640} \frac{m_{\rm i}}{\rho_{\rm i}} \left[\frac{\rho_{\rm i}}{\mu_{\rm i} m_{\rm H}}\right]^2 f_{\rm e} f_{\rm He III}\mbox{\ ,}
\end{equation}
where the $j$ are the temperature-dependent emission coefficients for the lines (Osterbrock \& Ferland 2006), and 
$f_{\rm He III}$ is the fraction of helium nuclei in He~{\sc iii} for each SPH particle.

Given the luminosity in a recombination line over an area of the sky, we may compute the flux in that line, as observed at $z$ = 0 with a spectral 
resolution $R$ = $\lambda$/$\Delta$$\lambda$, where $\lambda$ is the wavelength at which the emission line is observed (e.g. Oh et al. 2001).  
While Ly$\alpha$ photons are scattered out of the line of sight in the IGM prior to reionization (Loeb \& Rybicki 1999), a process which we do treat 
in the present calculations, H$\alpha$ and He~{\sc ii} $\lambda$1640 photons will not suffer such severe attenuation.  
Assuming that the line is unresolved, the monochromatic flux in H$\alpha$, for example, is

\begin{eqnarray}
f_{\rm H\alpha} &  = & \frac{l_{\rm H\alpha} \lambda_{\rm H\alpha}(1+z)R }{4\pi c D_{\rm L}^2(z)} \nonumber \\ 
                &  \sim & 20 {\rm\,nJy}\left(\frac{l_{\rm H\alpha}}{10^{40} {\rm erg\, s^{-1}}}\right)
                \left(\frac{1+z}{10}\right)^{-1}\left(\frac{R}{1000}\right) \mbox{\ ,} 
\end{eqnarray} 
where $l_{\rm H\alpha}$ is the luminosity in H${\rm \alpha}$ along the line of sight through the emitting galaxy, $D_{\rm L}(z)$ is the luminosity distance at redshift 
$z$ ($\sim$ 10$^2$ Gpc at $z$ = 10), and $\lambda_{\rm H\alpha}$ is the rest frame wavelength of the line, 656.3 nm.  If the galaxy is spatially unresolved, appearing as a point source, we may simply substitute $L_{\rm H\alpha}$ for $l_{\rm H\alpha}$ in equation~(10), to compute the total flux from the galaxy.  
In terms of total (integrated) line flux, we have the equivalent expression:

\begin{eqnarray}
F_{\rm H\alpha} &  = & \frac{L_{\rm H\alpha} }{4\pi D_{\rm L}^2(z)} \nonumber \\ 
                &  \sim & 10^{-20} {\rm erg\, s}^{-1} {\rm cm}^{-2} \left(\frac{L_{\rm H\alpha}}{10^{40} {\rm erg\, s^{-1}}}\right)
                \left(\frac{1+z}{10}\right)^{-2}\nonumber \mbox{\ .} 
\end{eqnarray}

\subsubsection{Recombination line equivalent widths}

Another observable quantity obtained from our simulations is the rest-frame equivalent width (EW) of recombination lines.  
We calculate the EWs of the three recombination lines considered, following Schaerer (2002):
\begin{equation}
W_{\rm H\alpha}^0 = \frac{L_{\rm H\alpha}}{L_{\lambda\rm , neb} + L_{\lambda\rm , *}} \mbox{\ }
\end{equation}

\begin{equation}
W_{\rm Ly\alpha}^0 = \frac{L_{\rm Ly\alpha}}{L_{\lambda\rm , neb} + L_{\lambda\rm , *}} \mbox{\ }
\end{equation}

\begin{equation}
W_{\rm \lambda 1640}^0 = \frac{L_{\rm \lambda 1640}}{L_{\lambda\rm , neb} + L_{\lambda\rm , *}} \mbox{\ ,}
\end{equation}
where the monochromatic continuum luminosity, evaluated at the wavelength of the line, is the sum of the nebular emission $L_{\lambda\rm , neb}$ and the 
stellar emission $L_{\lambda\rm , *}$.  The nebular continuum luminosity is given by

\begin{equation}
L_{\lambda\rm , neb} =  \frac{c}{\lambda^2} \frac{\gamma_{\rm tot}}{\alpha_{\rm B}} Q_{\rm rec}\mbox{\ ,}
\end{equation}
where $\lambda$ is the wavelength in question, and $Q_{\rm rec}$ is again the total 
number of recombinations per second in the halo.  The continuous emission coefficient $\gamma_{\rm tot}$ accounts for free-free, free-bound, and two-photon continuum emission, as described in Schaerer (2002).  The stellar continuum luminosity is calculated assuming a blackbody stellar spectrum and is given by
\begin{equation}
L_{\lambda\rm , *} = \frac{N_{\rm *}}{\lambda^5} \frac{8 \pi^2  h  c^2 R_{\rm *}^2}{{\rm exp}(hc / \lambda k_{\rm B} T_{\rm eff}) - 1}\mbox{\ ,}
\end{equation}
where $N_{\rm *}$ is the number of stars in the cluster, $T_{\rm eff}$ is the effective surface temperature of a star, and $R_{\rm *}$ is the stellar radius.

\section {Results and Implications}
We next discuss the observable characteristics
of primordial dwarf galaxies. In particular, 
we evaluate the utility of indicators for the SFR and the stellar IMF in such galaxies.

\begin{figure}
\vspace{2pt}
\epsfig{file=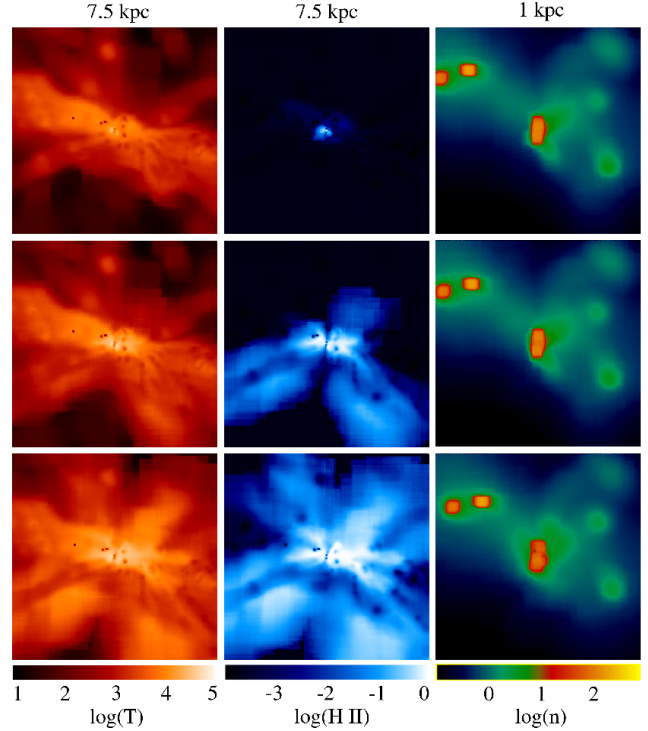,width=8.5cm,height=9.9cm}
\caption{The density-weighted temperature ({\it left column}), density-weighted H~{\sc ii} fraction ({\it middle column}), and number density ({\it right column}), 
each averaged along the line of sight, of the gas surrounding the more massive 100 ${\rmn M}_{\odot}$ star cluster, shown at three different times from the 
prompt formation of the cluster: 
500,000 yr ({\it top row}), 1 Myr ({\it middle row}), and 3 Myr ({\it bottom row}).  The H~{\sc ii} region grows as the density of the gas in the center 
of the host halo gradually drops in response to the intense photoheating.  Note the different length scale of each column, given at the top in physical units; the density is shown only within the central region of the host halo.}
\end{figure}

\begin{figure}
\vspace{2pt}
\epsfig{file=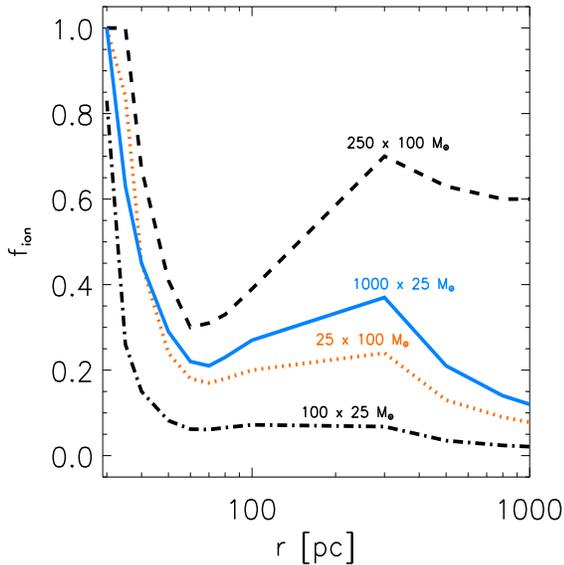,width=8.5cm,height=8.5cm}
\caption{The mass fraction of enclosed gas which is photoionized, $f_{\rm ion}$, as a function of the distance from the central star cluster, after 3 Myr 
of photoheating.  Each line corresponds to a different choice of IMF and total mass in stars, as labeled.  The ionized fraction begins to drop at $\geq$ 300 pc mostly
due to two minihaloes at $\sim$ 400 and $\sim$ 500 pc, shown in Fig.~2 (right panels), which remain shielded from the ionizing radiation 
and are thus largely neutral.  Within $\sim$ 100 pc, there is a large fraction of gas which remains un-ionized at densities $\geq$ 100 cm$^{-3}$.  This gas
will likely collapse to form stars, despite the strong radiative feedback from the central stellar cluster.}
\end{figure}

\subsection{Evolution of gas inside the galaxy}
With the ignition of a stellar cluster at the center of the host halo, the gas surrounding the cluster is photoheated, raising its pressure 
and leading to its outward expansion.  In turn, the overall recombination rate in the host halo drops, allowing the expansion of the H~{\sc ii} region
to continue for a constant rate of ionizing photon production. Fig.~2 shows the growth of the H~{\sc ii} region 
and the concomitant expansion of the gas in the center of the host halo for the more massive 100 ${\rmn M}_{\odot}$ cluster.  The H~{\sc ii} region breaks out of the host halo within the first $\sim$ 1 Myr, and after 3 Myr it extends to $\la$ 7 physical kpc, only slightly larger than the size of the H~{\sc ii} region created by a single massive Pop~III star in a minihalo 
(e.g. Alvarez et al. 2006).

While the gas surrounding the formation sites of the first stars in minihaloes
is easily photoevacuated by a single massive Pop~III star (e.g. Kitayama et al. 2004; Whalen et al. 2004), the deeper gravitational potential well of the 
DM haloes hosting the first galaxies allows for the retention of gas even under intense photoheating; indeed, this is one criterion used to define the first galaxies 
(e.g. Read 2006; Johnson et al. 2008; Greif et al. 2008).   

As shown in Fig. 3, a substantial portion of the gas in the galaxy, even within $\sim$ 100~pc of the stellar cluster, remains neutral after 3 Myr.  This gas
is shielded from the ionizing radiation, causing the ionization front (I-front) to propagate 
outward anisotropically in the inhomogeneous cosmological density field (see also e.g. Shapiro et al. 2004; Abel et al. 2006; Alvarez et al. 2006).  
Even for the case of the highest ionizing flux, the fraction of ionized gas within 
the central $\sim$ 100~pc is $\la$ 0.4, leaving the majority of the high density gas neutral.  While the photodissociating radiation from 
the initial stellar cluster will slow the collapse of this primordial gas (e.g. Susa \& Umemura 2006; Ahn \& Shapiro 2007; Whalen et al. 2008), 
some fraction of it will likely be converted into stars once the most massive stars in the cluster have died out. Indeed, the shocks engendered by the supernovae 
that mark the end of their lives may expedite the collapse of the gas (e.g. Mackey et al. 2003; Salvaterra et al. 2004; Machida et al. 2005; Greif et al. 2007; Sakuma \& Susa 2009).  
The incomplete ionization of the central gas confirms that the masses that we have chosen for the clusters are not overly large, 
as there is still some neutral gas available for subsequent star formation regardless of the radiative feedback.    

The gas that is photoionized, however, is gradually expelled from the center of the halo, and after 3 Myr of photoheating the density of the 
photoionized gas within $\sim$ 20 pc of the cluster drops to $\la$ 10 cm$^{-3}$ for the more massive cluster of 100 ${\rmn M}_{\odot}$ 
stars shown in Fig. 2.  
For our other choices of IMF and total cluster mass the dynamical response is less dramatic, as the ionizing flux is weaker; for example, 
after 3 Myr the density of the central
photoionized gas is $\la$ 50 cm$^{-3}$ for the less massive cluster of 25 ${\rmn M}_{\odot}$ stars.  The varying degree to which photoheating 
dynamically impacts the host halo leads to important differences in the properties of the emitted radiation.         

Although the limited resolution of our simulations allows only to track the expansion of the H~{\sc ii} region from an initial physical size of $\ga$ 10 pc,
we expect that after roughly a sound-crossing time of the central unresolved $\sim$ 10 pc, or after the first few 10$^5$ yr, the evolution of the 
H~{\sc ii} region is reliably resolved.  It should thus be noted that the breakout of the H~{\sc ii} region may be delayed by of order this 
timescale compared to our simulations.
We note that in the Milky Way the expansion of the photoheated gas in an H~{\sc ii} region may be slowed due to turbulent pressure confinement 
(e.g. Xie et al. 1996; Mac Low et al. 2007), resulting in ultra-compact H~{\sc ii} regions persisting for $\ga$ 10$^5$ yr (Wood \& Churchwell 1989), 
much longer than the sound-crossing timescale for such regions.  As turbulence begins to play an important role in the formation of the first galaxies 
(Wise et al. 2008; Greif et al. 2008), the initial evolution of H~{\sc ii} regions therein may be similarly confined.  
This possibility notwithstanding, we expect that the spatial resolution that we do achieve suffices to track changes in the luminosity emitted in 
recombination lines and in the escape fraction of ionizing radiation, which we discuss in the remainder of this Section.

\subsection{Star formation rate indicators}
The luminosity emitted in recombination lines, such as H$\alpha$, has been found to scale remarkably well with the SFR of galaxies at low redshift 
(e.g. Kennicutt 1983; but see Pflamm-Altenburg et al. 2007).  The SFR obtained using such relations relies on some knowledge of the IMF of the stars 
which are forming, as well as on the escape fraction of ionizing radiation. Fig.~4 shows our calculations of the escape fraction $f_{\rm esc HII}$ of 
H~{\sc i}-ionizing photons 
for each of our choices of IMF and total mass in stars, and Fig.~5 shows the escape fraction $f_{\rm esc HeIII}$ of He~{\sc ii}-ionizing photons. 
The corresponding luminosities emitted in H$\alpha$, Ly$\alpha$, and  
He~{\sc ii} $\lambda$1640 are presented in Fig.~6.                

As shown in Fig. 4, there is a clear trend toward higher H~{\sc i}-ionizing photon escape fractions for more massive stellar clusters, 
with the majority of the ionizing photons 
escaping from clusters with the larger total stellar mass.  The escape fraction is not, however, independent of IMF; for a given total mass in stars, 
the escape fraction can differ by a wide margin.  Both the variability in the escape fraction with time and the range of values that we find 
are in rough agreement with other recent calculations of the escape fraction of ionizing photons from dwarf galaxies at $z$ $\ga$ 10 (Wise \& Cen 2009; 
Razoumov \& Sommer-Larsen 2009).  For a recent calculation of the escape fraction of ionizing photons from more massive galaxies, see Gnedin et al. (2008). 

The breakout of the H~{\sc ii} region 
generated by the less massive 100 ${\rmn M}_{\odot}$ star cluster occurs after $\sim$ 1 Myr, leading to an escape fraction $\ga$ 0.5 after 2 Myr.  
In contrast, the H~{\sc ii} region of the equally massive 25 ${\rmn M}_{\odot}$ star cluster remains confined to the host halo for $\ge$ 3 Myr, contributing no 
ionizing photons to the IGM.  The progress of the initial stages of hydrogen reionization, likely driven by star formation in the first galaxies (e.g. Loeb 2008), 
may thus depend on whether these galaxies hosted massive ($\ga$ 10 ${\rmn M}_{\odot}$) or very massive stars ($\ga$ 100 ${\rmn M}_{\odot}$)(see also Choudhury \& Ferrara 2007).  

\begin{figure}
\vspace{2pt}
\epsfig{file=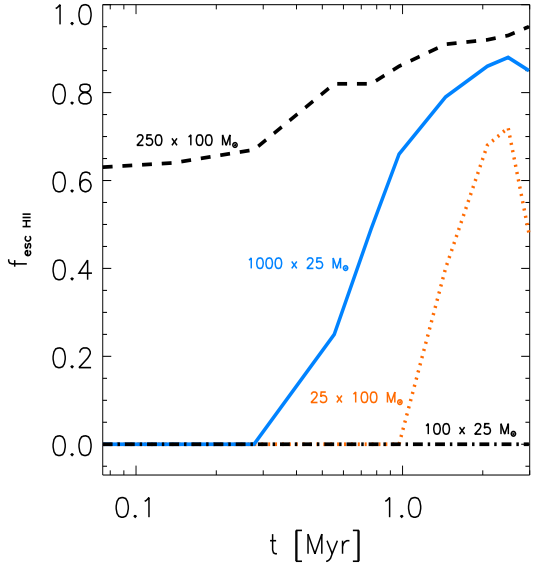,width=8.5cm,height=8.5cm}
\caption{The escape fraction of hydrogen-ionizing photons, $f_{\rm esc HII}$, from the host galaxy, each line corresponding to a different choice of IMF and total mass in stars, as labeled.  Note the tight anticorrelation between the escape fraction plotted here and the luminosity in the hydrogen recombination lines shown in Fig. 6, demonstrating that the vast majority of the energy emitted in hydrogen recombination lines emanates from the dense ionized gas within the host halo, as is shown in detail in Fig. 10.}
\end{figure}

\begin{figure}
\vspace{2pt}
\epsfig{file=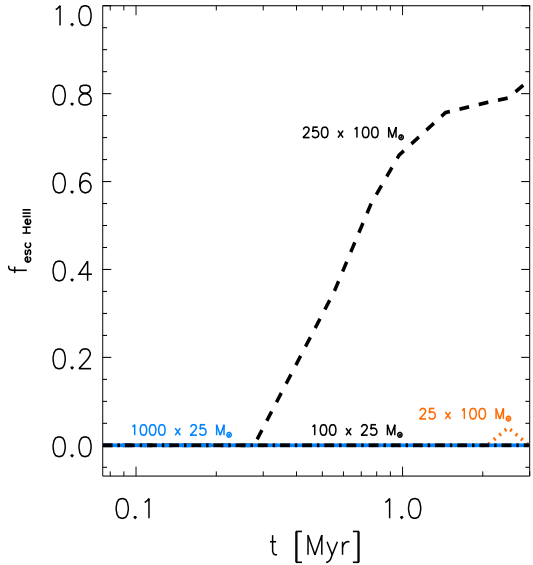,width=8.5cm,height=8.5cm}
\caption{The escape fraction of He~{\sc ii}-ionizing photons, $f_{\rm esc HeIII}$, from the host galaxy, each line labeled as in Fig. 4. For most cases, the negligible escape fraction leads to a tight correlation between the luminosity emitted in the He~{\sc ii} $\lambda$1640 line and the total mass contained in stars, in contrast to the weaker correlation for hydrogen recombination lines, as discussed in Section 3.2.}
\end{figure}

The evolution of the H~{\sc i}-ionizing photon  escape fraction is reflected in the evolution of the luminosity of hydrogen recombination lines, as shown 
in Fig. 6.  Comparing the panels 
on the left to those on the right, the luminosity in the Ly$\alpha$ and H$\alpha$ lines, while generally higher for larger total mass in stars, does not scale 
with the total mass in stars.  Indeed, owing to the increase in the escape fraction of ionizing photons, after $\sim$ 1 Myr the luminosity in hydrogen 
recombination lines from the clusters with greater total stellar mass drops below that of the clusters with lower total stellar mass, for a given IMF.  
Overall, because of the temporal evolution of the luminosity in a given line, there is no one-to-one relationship between the 
total mass in stars and the luminosity in a given recombination line.  There is thus 
likely to be a relatively 
weak correlation between the SFR and the luminosity in the hydrogen recombination lines emitted from the first dwarf galaxies, 
owing to the dynamical evolution of the photoionized gas and the escape of ionizing radiation into the IGM.

Similar to the case of the hydrogen recombination lines, the luminosity in the He~{\sc ii} $\lambda$1640 line is anticorrelated with the escape fraction of 
He~{\sc ii}-ionizing photons, shown in Fig. 5.  However, different from the case of the hydrogen lines, the luminosity emitted in He~{\sc ii} $\lambda$1640 line 
is generally much more strongly correlated with the total mass in stars, for a given IMF.  This is due to the low escape fraction of He~{\sc ii}-ionizing photons,
which is essentially zero for every case studied here, except for the case of the more massive cluster of 100 ${\rmn M}_{\odot}$ stars.  With such a high fraction of
He~{\sc ii}-ionizing photons being balanced by recombinations within the host halo, there is a near linear relation between the total mass in stars and the luminosity emitted in He~{\sc ii} $\lambda$1640, making this line a potentially much more reliable SFR indicator than hydrogen lines such as H$\alpha$.  There are slight departures 
from linearity due to the temperature dependence of the emission coefficient $j_{\rm \lambda 1640}$, which varies by a factor of $\sim$ 2 over the temperature range 
of the ionized gas in our simulations and is generally lower for the hotter H~{\sc ii} regions generated by the more massive stellar clusters 
(Osterbrock \& Ferland 2006).       

\begin{figure}
\vspace{2pt}
\epsfig{file=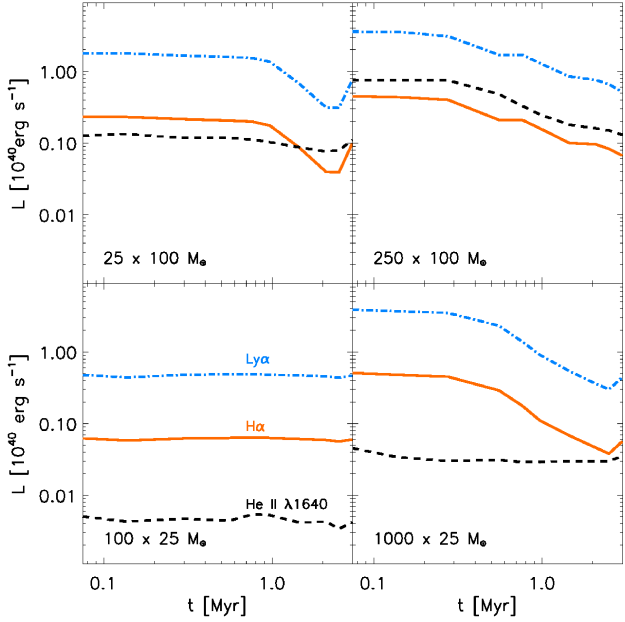,width=8.2cm,height=8.2cm}
\caption{The luminosity of the galaxy, as a function of the time from the prompt formation of the cluster, in three recombination lines: 
Ly$\alpha$ ({\it dot-dashed blue}), H$\alpha$ ({\it solid red}), and He~{\sc ii} $\lambda$1640 ({\it dashed black}).  
The four panels correspond to our four different choices of IMF and total mass in stars; these are, clockwise from top-left: twenty-five 100 
${\rmn M}_{\odot}$ stars, two hundred fifty 100 ${\rmn M}_{\odot}$ stars, one thousand 25 ${\rmn M}_{\odot}$ stars, and one hundred 25${\rmn M}_{\odot}$
stars.  The luminosities generally decrease with time, as the photoheating acts to decrease the density of the ionized gas, lowering 
the recombination rate.  Note the different evolution of the He~{\sc ii} $\lambda$1640 luminosity as compared to that of the hydrogen recombination lines, 
owing to the lower escape fraction of He~{\sc ii}-ionizing photons (see Fig. 5). 
}
\end{figure}

\begin{figure}
\vspace{2pt}
\epsfig{file=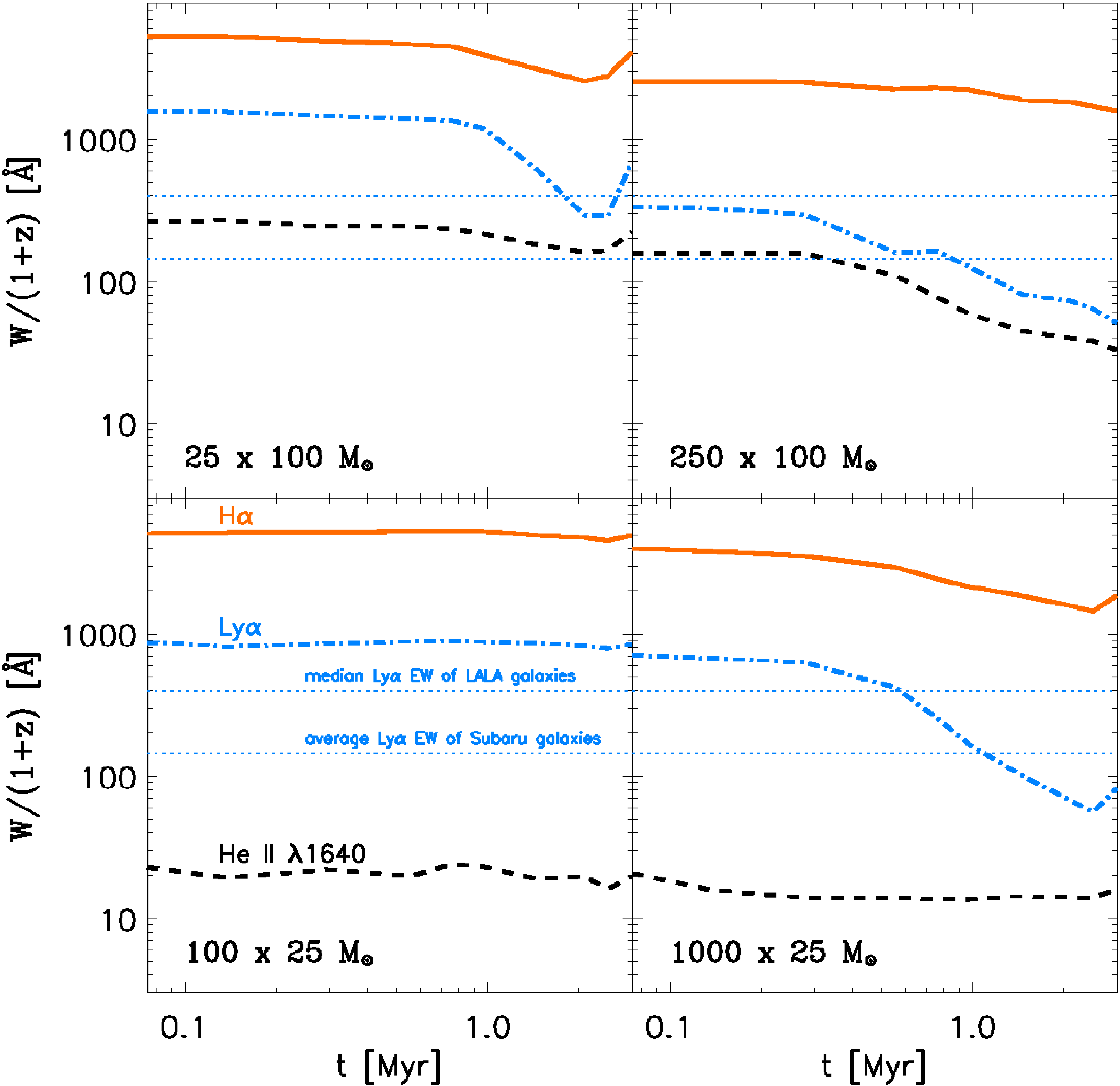,width=8.2cm,height=8.185cm}
\caption{The rest frame equivalent widths, $W^0$ = $W$ / (1+$z$), where $W$ is the observed EW,
as a function of time, of the same three recombination lines shown in Fig. 6.  For comparison, 
in each panel we plot the observed EWs of galaxies from two different surveys:  the {\it dotted line} 
at 400 $\AA$ denotes the median EW of Ly$\alpha$ emitters detected at $z$ = 4.5 in the LALA survey (Malhotra \& Rhoads 2002),
while the {\it dotted line} at 145 $\AA$ denotes the average EW of the six Ly$\alpha$ emitters at $z$ $\ge$ 6 detected 
in the Subaru deep field (Nagao et al. 2007).  Note that the Ly$\alpha$ EWs that we compute are upper limits, as scattering 
in a neutral IGM has not been accounted for.}
\end{figure}

\subsection{Initial mass function indicators}

The luminosity emitted from a galaxy in recombination lines depends not only on the stellar IMF, but also on the 
density field of the galaxy and the escape fraction of ionizing photons.  Therefore, the utility of recombination line strengths as
IMF indicators hinges on an understanding of the dynamical evolution of the photoionized gas, especially for the case of starbursts in the 
first dwarf galaxies, in which such dynamical effects can be most pronounced.  

For the starbursts that we simulate here, the luminosity of the He~{\sc ii} $\lambda$1640 emission line relative to the hydrogen recombination lines can be read 
from Fig. 6, while the equivalent widths of these lines are presented in Fig. 7.  Comparing the top panels of Fig. 6 to the bottom panels, it is evident that the 
ratio of the luminosity emitted in He~{\sc ii} $\lambda$1640 to that in H$\alpha$ (or Ly$\alpha$) can be very different depending on the IMF.  
Fig. 7 shows that there is a similar distinction in the ratios of the EWs.  For the 
100 ${\rmn M}_{\odot}$ star clusters the luminosity in He~{\sc ii} $\lambda$1640 is comparable to that in H$\alpha$, while for the 25 ${\rmn M}_{\odot}$ 
star clusters the luminosity in He~{\sc ii} $\lambda$1640 is up to an order of magnitude lower than that in H$\alpha$. However, as the escape fraction of 
H~{\sc i}-ionizing photons increases with time for the more massive 25 ${\rmn M}_{\odot}$ stellar cluster, the luminosities in these two lines 
become comparable, revealing that there is some ambiguity in the use of this ratio of line luminosities as an indicator of the IMF of young 
($\la$ 3 Myr) stellar clusters.  Thus, in some cases dynamical 
effects may compromise the use of this line ratio in distinguishing between clusters of Pop~III.1 and Pop~III.2 stars, with typical masses of order 100 
${\rmn M}_{\odot}$ and 10 ${\rmn M}_{\odot}$, respectively.

The ratio of the observed fluxes in He~{\sc ii} $\lambda$1640 and H$\alpha$, as calculated using equation (10), is displayed in Fig. 8.  In this Figure, it is 
clear that this line ratio is sensitive to the IMF, although it is not a constant for each cluster.  Instead, for clusters in which the 
escape fraction of H~{\sc I}-ionizing 
photons increases with time dramatically, while the escape fraction of He~{\sc II}-ionizing photons remains roughly constant, this line ratio varies 
with the flux observed in H$\alpha$.  While the ratio of the fluxes is a somewhat ambiguous IMF indicator, the clusters with the more top-heavy 
IMF do consistently exhibit larger ratios of He~{\sc ii} $\lambda$1640 to H$\alpha$.  Nagao et al. (2005) present a search for He~{\sc ii} $\lambda$1640 emission
from a strong Ly$\alpha$ emitter at $z$ = 6.33, finding an upper limit for the ratio of He~{\sc ii} $\lambda$1640 to Ly$\alpha$.  Assuming a standard value of 
0.07 for the ratio of luminosity in H$\alpha$ to that in Ly$\alpha$ (Osterbrock \& Ferland 2006), we show in Fig. 8 the upper limit that these authors report 
(see also Dawson et al. 2004).  Although a weak upper limit, it is clear that observations with only slightly greater sensitivity will allow to differentiate 
between the flux ratios predicted here for massive and very massive Pop~III IMFs.

Comparing the EW of H$\alpha$ in the four panels of Fig. 7, it is clear that it is not strongly dependent on the IMF or on the total mass in stars, 
varying by at most a factor of three between each of the cases.  While showing more variation between the four cases, the EW of Ly$\alpha$ also shows considerable 
ambiguity as an IMF indicator, its maximum value varying by about a factor of three between each of the four cases.  This insensitivity of the Ly$\alpha$ EW to 
the IMF arises from two effects.  Firstly, the stellar continuum luminosity $L_{\rm \lambda, *}$ increases in a similar manner as the number of ionizing photons 
from the massive (25 ${\rmn M}_{\odot}$) IMF to the very massive IMF (100 ${\rmn M}_{\odot}$).  This acts to keep the EW, roughly the ratio of the two, 
relatively constant.  Secondly, while the luminosity in Ly$\alpha$ decreases with the increasing escape fraction of ionizing photons for the more massive clusters,
the continuum luminosity remains largely unchanged, leading to a decrease in the EW with time for these clusters.  We note that the Ly$\alpha$ EWs presented here are only upper limits, as we have not accounted for scattering of Ly$\alpha$ photons in the IGM (see e.g. Dijkstra et al. 2007).  

The EW of He~{\sc ii} $\lambda$1640 is a more definitive indicator of IMF, being higher for the clusters of 100 ${\rmn M}_{\odot}$ stars than for the 
clusters of 25 ${\rmn M}_{\odot}$ stars, regardless of the total mass in stars or of the age of the cluster (up to $\ga$ 3 Myr).  
As with the utility of He~{\sc ii} $\lambda$1640 
as a SFR indicator, this largely follows from the generally low escape fraction of He~{\sc ii}-ionizing photons.    
\begin{figure}
\vspace{2pt}
\epsfig{file=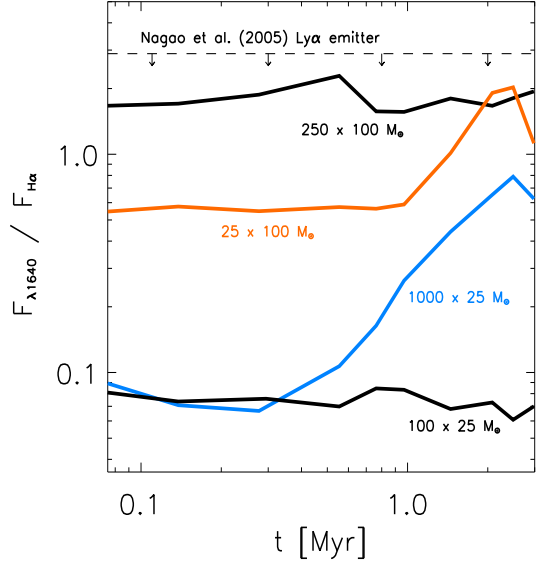,width=8.5cm,height=8.5cm}
\caption{The ratio of the integrated fluxes in He~{\sc ii} $\lambda$1640 and H$\alpha$, F$_{\rm \lambda 1640}$/F$_{\rm H\alpha}$, as a function of time, for each of the four clusters simulated here, as labeled.  The dashed horizontal line denotes the upper limit of this ratio for the strong Ly$\alpha$ emitter SDF J132440.6+273607 at $z$ = 6.33, as reported by Nagao et al. (2005).  Similar upper limits for Ly$\alpha$ emitters at $z$ = 4.5 have been reported by Dawson et al. 2004.}   
\end{figure}

For comparison with observed galaxies, we plot in Fig. 7 the two observational results: the median Ly$\alpha$ EW of galaxies detected in the Large Area Lyman Alpha (LALA) survey, $W^0 \sim 400 {\rm \AA}$, and the average EW of six galaxies observed at $z$ $\ge$ 6 in the Subaru deep field, $\sim 145 {\rm \AA}$ (Nagao et al. 2007).   
The large LALA EWs are comparable to what we find for primordial dwarf galaxies, although the LALA galaxies likely do not host Pop~III star formation 
(but see Jimenez \& Haiman 2006).  The detection of an EW of the He~{\sc ii} $\lambda$1640 line $\ga 10 {\rm \AA}$ would be a stronger indication of a 
galaxy hosting primordial star formation, as shown in Fig. 7, although none has been found as of yet.  We note that observed Lyman break galaxies at 
$z$ $\sim$ 3 have been found to have He~{\sc ii} $\lambda$1640 EWs of $\sim 2 {\rm \AA}$ (Shapley et al. 2003), consistent with what is expected for 
Wolf-Rayet stars formed in starbursts (see Schaerer \& Vacca 1998; Brinchmann et al. 2008).  
  
\begin{figure}
\vspace{2pt}
\epsfig{file=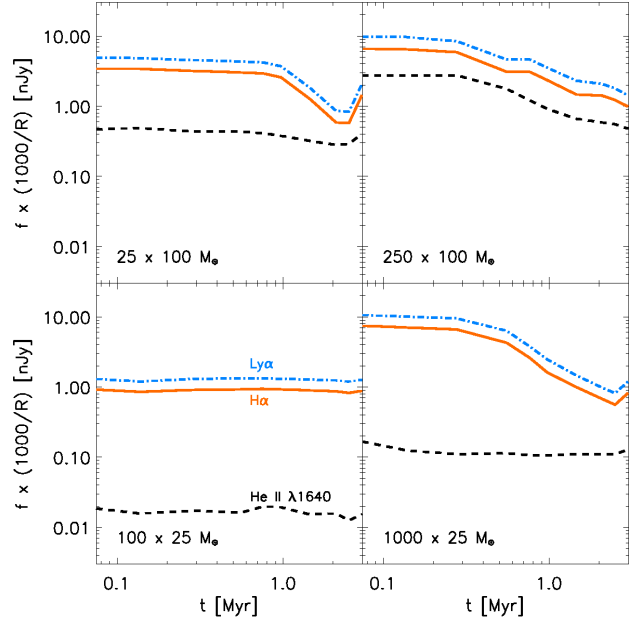,width=8.2cm,height=8.2cm}
\caption{The observed fluxes, $f$, as a function of time, of the same three recombination lines shown in Fig. 6, for the galaxy we simulate at $z$ $\sim$ 12.5.  The fluxes are normalized to what would be observed with a spectroscopic resolution of $R$ = 1000, and are computed using equation (10) assuming that the galaxy appears as an unresolved point source.  Note that the flux in Ly$\alpha$ is an upper limit, as the present calculation does not take into account scattering in a neutral IGM.}
\end{figure}

\subsection{Detectability of Recombination Radiation}  
In Fig.~9, we present our predictions for the observable recombination line fluxes, for each of the stellar clusters that we simulate.
Fig.~10 shows the surface brightness in H$\alpha$ as observed on the sky, for the two more massive stellar clusters, which each have a total 
mass in stars of 2.5 $\times$ 10$^4$ ${\rmn M}_{\odot}$.  
The fluxes in each plot, largely determined by our choices for the total stellar mass, are calculated using equation (10).
While the larger H~{\sc ii} region generated by the more massive stars encompasses more dense gas, creating more widely distributed emission in H$\alpha$, 
as shown in Fig. 10, the highest flux per square arcsecond is in the 
central region of the halo hosting the less massive stars.  This is due again to the less dramatic dynamical response of the gas to photoheating, 
leading to higher densities, and thus higher recombination rates.  
Due to this effect, the highest fluxes are generated just after the birth of a stellar cluster, as shown in Figure 9, 
when the density of 
the photoionized gas is still high, not having had time to expand in response to the concomitant heating.  Indeed, Fig. 9 shows that the flux in 
H$\alpha$ from the more massive 25 ${\rmn M}_{\odot}$
star cluster may reach $\ga$ 10 $\times$ ($R$/1000) nJy before the breakout of the H~{\sc ii} region.  Catching the first galaxies when still in the 
earliest stages of their initial starbursts, within the first few 10$^5$ yr, is thus likely to provide one of the best chances for observations of purely 
primordial stellar populations in the early Universe.    

\begin{figure*}
\includegraphics[width=6.95in]{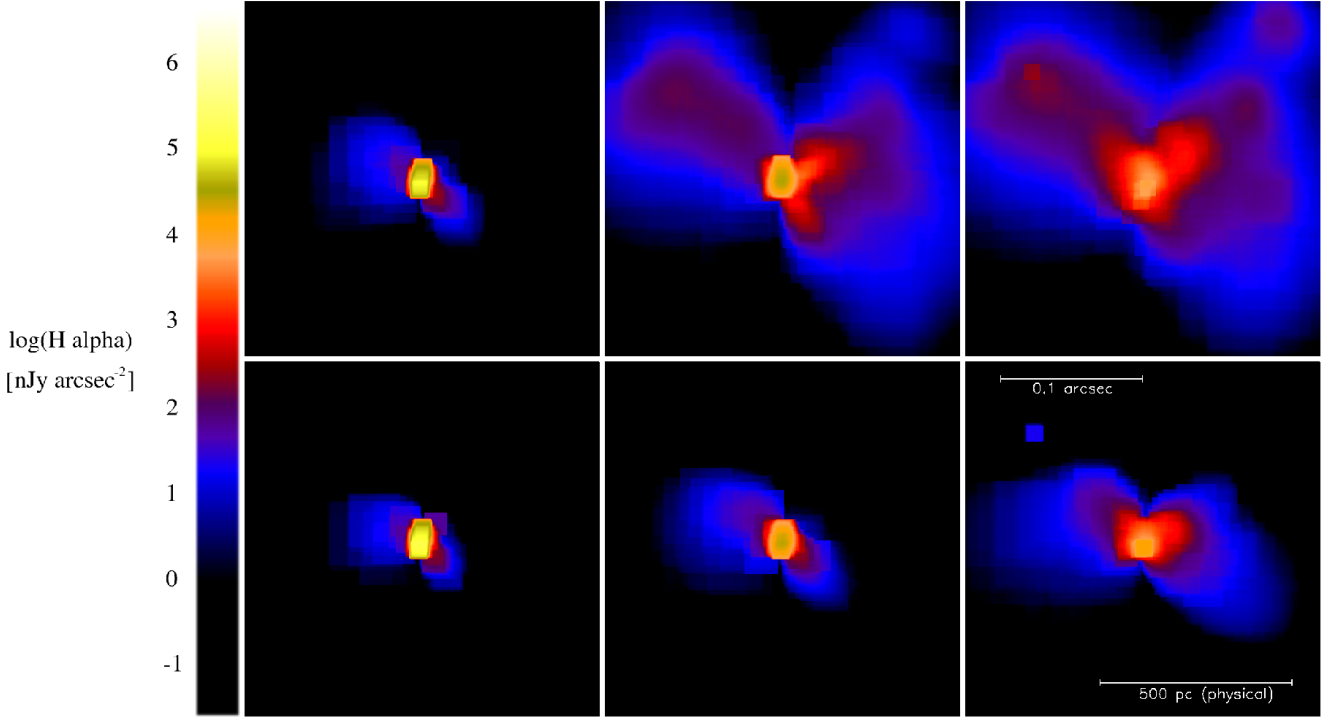}
\caption{The flux in H$\alpha$, $f_{\rm H\alpha}$, per square arcsecond, emitted from a primordial dwarf galaxy, as observed on the sky at $z$ = 0, 
assuming a spectroscopic resolution of R = 1000.  Shown here are the two most massive of the four 
stellar clusters that we simulate, one containing 25 ${\rmn M}_{\odot}$ stars ({\it bottom panels}), the other 
containing 100 ${\rmn M}_{\odot}$ stars ({\it top panels}).  From left to right, the clusters are shown at 10$^5$ yr, 1 Myr, and 3 Myr after
formation. 
 Note that the emission is concentrated in the densest photoionized regions, the filaments around the galaxy (compare the top panels to Fig. 2) 
and especially the dense gas within the inner $\sim$ 100 pc of the galaxy.  The highest total fluxes occur at the earliest times, before the H~{\sc ii} region has broken out; hence, the youngest stellar clusters are the most readily observed. 
}
\end{figure*}
  
Aboard the JWST, the H$\alpha$ line would be observed with the Mid-Infrared Instrument (MIRI). Its pixel size of $\ga$ 0.1 arcsec 
would not resolve the brightest portions of the galaxies that we simulate, which, as shown in Fig. 10, are of order 0.01 arcsec.  With a resolution capability 
of $R$ = 3000 the MIRI has a sensitivity of $\ga$ 200 nJy for a signal-to-noise of 10, in exposures of $\sim$ 10$^6$ s (see Panagia 2004), making it unable to detect even the brightest galaxies that we simulate, the flux in H$\alpha$ of these being $\la$ 20 nJy for $R$ = 3000.  

With a greater sensitivity of $\la$ 100 nJy (Panagia 2004), the Near Infrared Spectrograph (NIRSpec) operates in the wavelength range 0.7 to 5 $\mu$m, allowing it 
to possibly detect Ly$\alpha$ out to $z$ $\sim$ 40 and He~{\sc ii} $\lambda$1640 out to $z$ $\sim$ 30.  However, the flux in He~{\sc ii} $\lambda$1640 is 
always below that in H$\alpha$ and, hence, is too low to be detected.  
The Ly$\alpha$ line, with the highest flux of the three recombination lines shown in Fig. 9, is also not directly detectable, with a flux falling well below the $\sim$ 
100 nJy sensitivity limit of NIRSpec. Furthermore, although the 
luminosity in Ly$\alpha$ is always intrinsically higher than that in H$\alpha$, before reionization the observable flux in Ly$\alpha$ may be dramatically 
decreased due to scattering in the neutral IGM (e.g. Dijkstra et al. 2007). Because we do not account for this effect in the present calculations, the Ly$\alpha$ fluxes presented here are only upper limits.

The Near Infrared Camera (NIRCam) aboard the JWST, which will be used to conduct deep surveys designed to detect the first galaxies, will be capable of 
detecting point source fluxes as low as $\sim$ 3.5 nJy at a signal-to-noise of 10 for a 10$^5$ second exposure (e.g. Gardner et al. 2006).  With a resolution of 
$\ga$ 0.03 arcsec per pixel, the NIRCam would also not quite resolve the galaxies that we simulate.  We can evaluate the possibility that NIRCam may detect them as 
point sources, however, by estimating the continuum flux of the galaxies as observed at $\sim$ 2 $\mu$m.  As can be read from Figs. 6 and 7, 
the continuum flux, $\propto$ $L_{\rm \lambda1640}$ / $W_{\rm \lambda1640}$, varies only by a factor of a few between Ly$\alpha$ and He~{\sc ii} $\lambda$1640.  Thus,
for simplicity we assume that the continuum is roughly flat and calculate the specific continuum flux, as observed at $z$ = 0, as 

\begin{eqnarray}
f_{\rm cont} & \sim & \frac{L_{\rm \lambda1640}}{4 \pi c D_{\rm L}^2(z)} \frac{\lambda_{\rm \lambda1640}^2 (1+z)}{W_{\rm \lambda1640}^0 } \nonumber \\
             & \sim & 0.03 {\rm nJy}                            \mbox{\ ,} 
\end{eqnarray}
where $W_{\rm \lambda1640}^0$ is the equivalent width in the rest frame of the galaxy, as defined in equation (9).  This flux is well below the sensitivity limit 
of the NIRCam, and so we conclude that detection of the continuum radiation from the galaxies we simulate would also be undetectable.  
We note, however, that 
under favorable circumstances, gravitationally lensed emission from a primordial galaxy undergoing an only slightly more luminous starburst may be detectable with the JWST, 
given that lensing can boost the flux by a factor of order 10 (e.g. Refsdal 1964; Stark et al. 2007).

The first dwarf galaxies could be more luminous than we find here if the efficiency of star formation $\epsilon_{\rm SF}$, defined as the fraction of the total 
baryonic mass in 
the galaxy contained in stars, is larger than what we have assumed in our simulations, where our choices for the total mass in stars correspond to modest values of $\epsilon_{\rm SF}\sim 
10^{-3} - 10^{-4}$. A larger efficiency ($\epsilon_{\rm SF}$ $\sim$ 10$^{-1}$), for the top-heavy IMFs considered here, would yield a cluster
observable by NIRCam (e.g. Gardner et al. 2006).  However, as we have demonstrated, the much higher ionizing flux from a $\sim$ 10$^6$  
${\rmn M}_{\odot}$ cluster
of massive primordial stars would induce a strong hydrodynamic response which would lead to a rapid decline in the luminosity emitted in 
recombination radiation.  Thus, even if such clusters can be identified by their continuum emission, the detection of recombination radiation, and 
with it information about the stellar IMF, may be beyond the capabilities of the JWST.  Furthermore, the formation of such a massive cluster 
of primordial stars may face an impediment due to the strong radiative feedback within the cluster itself.  Recent simulations of 
clustered star formation in the present-day Universe suggest that radiative feedback influences the fragmentation behavior of the gas and possibly lowers 
the overall star formation efficiency (see Krumholz et al. 2007; Bate 2009).  However, the situation is by no means clear (e.g. Dale et al. 2005, 2007).  

It is possible that more massive (10$^9$ - 10$^{10}$ ${\rmn M}_{\odot}$) primordial galaxies form at $z$ $\ga$ 12, or form at lower redshift, making their detection feasible. However, such more massive, and therefore more luminous, galaxies are 
likely to also be more chemically evolved, and so may already be dominated by Pop~II star formation.  Thus, it may be that the galaxies 
which host pure Pop~III starbursts, such as those we study here, will remain out of reach of even the JWST, although this critically 
depends on the poorly constrained process of metal enrichment in the early Universe 
(see e.g. Pan \& Scalo 2007; Tornatore et al. 2007; Cen \& Riquelme 2008; Johnson et al. 2008).  We emphasize, however, that the dynamical effects
studied here are likely to play a role even in the more luminous galaxies that will be detected, and are important to take into account in evaluating 
observations meant to constrain the SFR or the IMF.

\section {Summary and Conclusions}
We have presented calculations of the properties of the recombination radiation emitted from a primordial dwarf galaxy at $z$ $\sim$ 12.5, 
during the initial stages of a starburst. Our cosmological radiation-hydrodynamical simulations allow us to track the detailed dynamical 
evolution of the emitting gas in the central regions of the galaxy, and thus to study its effect on the emerging radiation.  The goal of this study has
been to determine the observable signatures of the initial starbursts in the first galaxies.  In particular, we have aimed to find reliable
indicators of the star formation rate and of the stellar IMF.   

Owing to the escape of H~{\sc i}-ionizing photons into the IGM, we find only a weak correlation between the total mass in stars and 
the luminosity in hydrogen recombination lines.  This suggests that Ly$\alpha$ and H$\alpha$, despite the high luminosity in these lines, may not serve 
as strong indicators of the SFR, unlike in the low-redshift Universe (e.g. Kennicutt 1983).  The He~{\sc ii} $\lambda$1640 line may be a more effective
SFR indicator, as the luminosity in this line scales more closely with the total mass in stars, due to the lower escape fraction of He~{\sc ii}-ionizing 
photons.     

We confirm that the ratio of He~{\sc ii} $\lambda$1640 to either Ly$\alpha$ or H$\alpha$ can be used as an IMF indicator, although
its utility is compromised in some cases by the unequal escape fractions of H~{\sc i}- and He~{\sc ii}-ionizing photons.  The most robust IMF indicator,  
in terms of distinguishing between populations of massive ($\ga$ 10 ${\rmn M}_{\odot}$) and very massive ($\ga$ 100 ${\rmn M}_{\odot}$) Pop~III stars, is the 
EW of  He~{\sc ii} $\lambda$1640, as it is consistently higher for the more massive stars regardless of the total mass in stars.  
We note that while in principle the radiation emitted by a central accreting black hole (BH) in a primordial dwarf galaxy could introduce complications 
for using He~{\sc ii} $\lambda$1640 as IMF indicator (e.g. Tumlinson et al. 2001), recent work suggests that BH accretion is inefficient 
in the early Universe (e.g. Johnson et al. 2007; Pelupessy et al. 2007; Alvarez et al. 2008; Milosavljevi\'c et al. 2008, 2009). Such miniquasar activity may 
thus not result in appreciable observable radiation for the first $\sim 10^8$~yr.

In cases where the gas in primordial dwarf galaxies is unable to cool efficiently due to the photodissociation of H$_{\rm 2}$, it has been argued that, 
instead of a stellar cluster, a BH with mass $\ga$ 10$^4$ ${\rmn M}_{\odot}$ may form by direct collapse (see Bromm \& Loeb 2003; 
Lodato \& Natarajan 2006; Begelman et al. 2006; Spaans \& Silk 2006; Regan \& Haehnelt 2008).  The fraction of dwarf galaxies 
that host the formation of such BHs is likely quite small, 
since the photodissociating flux required to suppress fragmentation is very high, well above the cosmological 
average at $z$ $\ga$ 10 (Dijkstra et al. 2008).  In the present work, we have chosen to focus on the presumably more common case 
that a stellar cluster forms (but see Begelman \& Shlosman 2009).  We do note, however, that stellar mergers within dense clusters, such as those we consider here, 
may lead to the formation of $\sim$ 10$^3$ ${\rmn M}_{\odot}$ BHs (Omukai et al. 2008; Devecchi \& Volonteri 2009).  

In terms of the detectability of the recombination radiation from the first galaxies, we have shown that due to the dynamical response of the gas 
to photoheating, a top-heavy IMF or a high star formation efficiency can be self-defeating, leading to a decrease in the line luminosity of the galaxy.  
We conclude that the detection of purely primordial dwarf galaxies at $z$ $\ga$ 10 is likely to be beyond the capabilities of the JWST, although
their detection may be just possible if the galaxies are strongly lensed. More luminous, 
$10^9 - 10^{10} {\rmn M}_{\odot}$ (total mass), galaxies may thus be detected by the JWST; however, such more massive galaxies are likely to be already chemically enriched due to previous episodes of star formation in their progenitors. 
These systems are then expected to host Pop~II star formation, or a composite
of Pop~II and Pop~III, depending on the still poorly understood mixing of
heavy elements in the first galaxies (e.g. Karlsson et al. 2008).

Our results demonstrate how the radiation emitted from the first galaxies depends on the hydrodynamic effects of the photoionization from
clusters of massive stars.  This is complementary to the results of previous studies (e.g. Schaerer 2002, 2003), 
which highlight the evolution of the emitted radiation owing to the aging of a stellar population.  Clearly, both 
effects must be considered in future work.

The initial starbursts of the first galaxies may constitute the formation sites of the only metal-free stellar clusters in the Universe, 
since after the first several Myr supernova feedback can quickly enrich the galaxy with metals (e.g. Mori et al. 2002; Kitayama \& Yoshida 2005; 
but see Tornatore et al. 2007, Cen \& Riquelme 2008).  Also, a large fraction of the first dwarf galaxies, with masses of order 10$^8$ ${\rmn M}_{\odot}$, 
may already form from metal-enriched gas (Johnson et al. 2008; see also Omukai et al. 2008); it is an important open 
question what fraction of dwarf galaxies forming at $z$ $\ga$ 10 are primordial when they begin forming stars.  Future observations of those 
first dwarf galaxies that do host primordial star formation offer one of the few opportunities for constraining the primordial IMF.

\section*{Acknowledgments}
We would like to thank Josh Adams, Marcelo Alvarez, and Mark Dijkstra for helpful discussions. 
We are also grateful to Lars Hernquist for his comments on a previous version of this paper, as well as to 
the anonymous referee whose suggestions greatly improved the presentation of this work.
VB acknowledges support from NSF grant AST-0708795 and NASA ATFP grant
NNX08AL43G. THG thanks for travel support from the Heidelberg 
Graduate School of Fundamental Physics, funded by the Excellence Initiative of 
the German Government (grant number GSC 129/1). RSK thanks the German Science 
Foundation (DFG) for support via the Emmy Noether grant KL 1358/1 and also 
acknowledges subsidies from the DFG SFB 439 Galaxies in the Early Universe as 
well as from the FRONTIER program of Heidelberg University.
The simulations presented here were carried out at
the Texas Advanced Computing Center (TACC).

\end{document}